\documentstyle[12pt]{article}
\renewcommand{\theequation}{\thesection.\arabic{equation}}
\newcommand{\beq}{\begin{equation}}
\newcommand{\eeq}{\end{equation}}
\newcommand{\bea}{\begin{eqnarray}}
\newcommand{\eea}{\end{eqnarray}}

\begin{document}
\setcounter{page}{0}
\topmargin 0pt
\oddsidemargin 5mm
\renewcommand{\thefootnote}{\fnsymbol{footnote}}
\newpage
\setcounter{page}{0}
\begin{titlepage}
\begin{flushright}
QMW 93-19
\end{flushright}
\begin{flushright}
hep-th/9307163
\end{flushright}
\vspace{0.5cm}
\begin{center}
{\large {\bf Conformal Non-Abelian Thirring Models}} \\
\vspace{1.8cm}
\vspace{0.5cm}
{\large Oleg A. Soloviev
\footnote{e-mail: soloviev@V1.PH.QMW.ac.uk}\footnote{Work supported by
S.E.R.C.}
 }\\
\vspace{0.5cm}
{\em Physics Department, Queen Mary and Westfield College, \\
Mile End Road, London E1 4NS, United Kingdom}\\
\vspace{0.5cm}
\renewcommand{\thefootnote}{\arabic{footnote}}
\setcounter{footnote}{0}
\begin{abstract}
{The Lie-Poisson structure of non-Abelian Thirring
models is discussed and the Hamiltonian quantization of these theories
is carried out. The consistency of the Hamiltonian quantization with the path
integral method is established.
It is shown that the space of non-Abelian Thirring models contains the
nonperturbative
conformal points which are in one-to-one correspondence with general solutions
of the Virasoro master equation. A BRST nature of the mastert equation is
clarified.}
\end{abstract}
\vspace{0.5cm}
\centerline{June 1993}
\vspace{0.5cm}
\end{center}
\end{titlepage}
\newpage
\section{Introduction}

Thirring models appear to be of a great interest in string theory. This
interest has mainly come from the idea that the space of all conformal
Thirring models is the space of string compactifications. The Abelian
Fermionic Thirring models have been considered as the most appropriate
candidates to describe all toroidal string compactifications [1]. In their
turn the non-Abelian (Lefton-Righton) Thirring models have been proposed
for the description of the most general (left-right asymmetric) string
compactifications on group
manifolds [2, 3]. Therefore, the space of all conformal Thirring models seems
to
collect all symmetric string vacua, which could form the multitude of conformal
backgrounds appropriate to the formulation of background independent closed
string field theory [4].

The natural coordinates in the Thirring model-space might be the coupling
constants  of the current-current interaction. The whole model-space,
however, may have additional dimensions parameterized by some extra variables
coming from the geometric formulation of the Thirring model [5, 6]. Therefore,
it would be illuminating if one could explore the Thirring model at all the
possible values of its couplings. However, this seems to be beyond our present
analytical abilities. Most of the difficulty resides in the highly non-linear
character of the current-current interaction of the Thirring theory. Given our
present knowledge, the theory is tractable only when it possesses either affine
symmetry or quantum group symmetry (which might turn out to be a sort of
deformation of the former.) In this paper we will not discuss the quantum group
symmetry of Thirring models but rather affine symmetries. We will show that
affine symmetries are intimately related to the conformal invariance of the
Thirring model.

We will begin with a description of the two formulations of the non-Abelian
Thirring model. Namely, Fermionic and Bosonic formulations. We will show when
these two classically distinguished theories become equivalent at the quantum
level. In sect. 3 we will discuss the Lie-Poisson structure of the classical
Bosonic Thirring model. We will demonstrate the interconnection between this
algebraic structure of the Thirring model and self-duality of its
two-dimensional fields valued in  Lie algebras. Sect. 4 contains the
Hamiltonian quantization of the non-Abelian Thirring model on the basis of the
Lie-Poisson structure of the classical theory. It will be shown that the given
quantization will be consistent as long as the conformal symmetry is present.
We will find that consistency requires particular values of the
Thirring coupling constants
which are in one-to-one correspondence with solutions of the so-called Virasoro
master equation [7]. The Virasoro master equation describes in conformal field
theory the most general embedding of the Virasoro algebra into the affine
algebra through the affine-Virasoro construction [7]. The
affine-Virasoro construction in its turn is the most general bilinear of the
affine currents. Thus, the non-Abelian Thirring models provide a natural
sigma model interpretation to conformal theories based on general solutions of
the master equation.\footnote{Note that our field-theoretic realization of the
affine-Virasoro construction [6,10,11] differs from attempts in [8,9].} It is
interesting that the so-called Sugawara solution of the Virasoro master
equation has
been derived by Dashen and Frishman from the isoscalar
non-Abelian Thirring model two decades ago [12]. Therefore we call
all conformal points, which are solutions of the master equation,
Dashen-Frishman
conformal points [6,10,11]\footnote{In addition to the Dashen-Frishman
conformal
points non-Abelian Thirring models may have so-called ``Higgs conformal
points" [13] which are descendants of Dashen-Frishman conformal
points. Higgs conformal points come into being due to duality symmetry which
Thirring models possess at the quantum level [2, 13].}.
In sect. 5 we will derive these same Dashen-Frishman
conformal points from the Fermionic Thirring model. In the process we will
justify the
conformal symmetry at the Dashen-Frishman conformal points by using
the path integral method. In sect. 6 we will show that the Dashen-Frishman
conformal points appear to be a consistence condition for the BRST quantization
of the Bosonic-Thirring model coupling to the two dimensional gravity.
Finally, sect. 7 contains some concluding remarks.

In the appendix, we will discuss the representation of the affine-Virasoro
construction for the affine group $SU(2)$. The $SU(2)$ case is interesting
because the minimal conformal series can be described with the $SU(2)$
non-Abelian Fermionic Thirring models [11].

\section{Fermionic and Bosonic non-Abelian Thirring models}

Let us start with a description of the Fermionic non-Abelian Thirring model.
The
action is given by
\begin{equation}
S_F={1\over4\pi}
\int d^2z(\bar\psi_L\bar\partial\psi_L+\bar\psi_R\partial\psi_R-S_{a\bar a}
J_L^aJ_R^{\bar a}),\end{equation}
where $\psi_L$ and $\psi_R$ are complex Weyl spinors (in general carrying a
flavor) transforming as the fundamental representations of given groups $G_L$
and $G_R$ respectively. The last term in (2.1) describes the general
interaction
between fermionic currents
\begin{equation}
J^a_L=\bar\psi_L t^a\psi_L,\;\;\;\;\;J_R^{\bar a}=\bar\psi_Rt^{\bar a}\psi_R.
\end{equation}
Here $t^a,\;t^{\bar a}$ are the generators in the Lie algebras ${\cal G}_L,\;
{\cal G}_R$.
\begin{eqnarray}
[t^a,t^b]&=&if^{ab}_ct^c,\;\;\;a,b,c,=1,2,...,\dim G_L,\nonumber\\ & & \\
\left[t^{\bar a},t^{\bar b}\right]&=&if^{\bar a\bar b}_{\bar c}t^{\bar
c},\;\;\;
\bar a,\bar b,\bar c=1,2,...,\dim G_R.\nonumber\end{eqnarray}
$S_{a\bar a}$ is a coupling constant matrix.

We have used also the following notations
\begin{eqnarray}
&\partial \equiv\partial /\partial z,\;\;\;&\bar\partial
\equiv\partial /\partial
\bar z,\nonumber\\
&z=(t^E+ix)/{\sqrt2},\;\;\;&\bar z=(t^E-ix)/{\sqrt2},\nonumber\end{eqnarray}
where $t^E=it$. We will use $z$ and $\bar z$ to denote Euclidean coordinates,
whereas $x$ and $t$ will signify Minkowski coordinates. We follow the
convention
$d^2z\equiv idzd\bar z=-dxdt^E/2$.

In addition to the non-Abelian interaction, we always can include the $U(1)$
current-current interaction in the Fermionic action. The $U(1)$ currents are
defined as
\begin{equation}
J_L=\bar\psi_L\psi_L,\;\;\;\;J_R=\bar\psi_R\psi_R,\end{equation}
where a sum over all the internal indices is assumed.

It is worth mentioning that the classical Fermionic non-Abelian Thirring model
obviously possesses the global $G_L\times G_R$ invariance provided the
coupling matrix $S_{a\bar a}$ also transforms as the adjoint representation of
the $G_L\times G_R$ group. Therefore, the physically distinguished couplings
are
defined as
\begin{equation}
\tilde S_{a\bar a}=\{S_{a\bar a}\}/{\rm ad}(G_L\times G_R),\end{equation}
where $\{S_{a\bar a}\}$ is a set of all the consistent values of the Thirring
couplings $S_{a\bar a}$. However at the quantum level the given symmetry can be
broken or reduced to a smaller one. We will show that the exact symmetry of the
conformal points is the diagonal subgroup of the group $G_L\times G_R$. Due to
this symmetry, the space of Thirring models ought perhaps to be a coset.

The action of the Bosonic Thirring model is formulated as follows
\begin{equation}
S_B=\int\left[ L_L(k_L,g_L)+L_R(k_R,g_R)+L_{int}(g_L,g_R;S)\right],
\end{equation}
where these three terms respectively are given by
\begin{eqnarray}
4\pi L_L(k_L,g_L)&=&-k_L\left[(1/2){\rm tr}_L|g^{-1}_L{\rm d}g_L|^2+(i/3){\rm
d}
^{-1}{\rm tr}_L(g^{-1}_L{\rm d}g_L)^3\right],\nonumber\\
4\pi L_R(k_R,g_R)&=&-k_R\left[(1/2){\rm tr}_R|g^{-1}_R{\rm d}g_R|^2+(i/3){\rm
d}
^{-1}{\rm tr}_R(g^{-1}_R{\rm d}g_R)^3\right],\\
L_{int}(g_L,g_R;S)&=&-(k_Lk_R/4\pi){\rm tr}_L{\rm tr}_Rg^{-1}_L\partial g_L\;
S\;\bar\partial g_Rg^{-1}_Rdzd\bar z,\nonumber\end{eqnarray}
with the coupling $S$ belonging to the direct product ${\cal G}_L\otimes{\cal
G}_R$. Here the fields $g_L$ and $g_R$ take their values in the Lie groups
$G_L$ and $G_R$, respectively. $k_L,\;k_R$ are central elements in the affine
algebras $\hat{\cal G}_L,\;\hat{\cal G}_R$. The symbols ${\rm tr}_L,\;{\rm
tr}_R$ indicate tracing over the group indices of $G_L,\;G_R$.

The point to be made is that the non-Abelian Bosonic Thirring model in eq.
(2.6)
becomes equivalent to the Lefton-Righton Thirring model [13] when the fields
$g_L,\;g_R$ obey the conditions
\begin{eqnarray}
\partial_+g_Rg^{-1}_R&=&- k_L{\rm tr}_LS\;g^{-1}_L\partial_+g_L,\nonumber
\\ & & \\
g^{-1}_L\partial_-g_L&=&-k_R{\rm tr}_RS\;\partial_-g_Rg^{-1}_R,\nonumber
\end{eqnarray}
where we have used the light cone coordinates
\begin{eqnarray}
&x^+=x+t,\;\;\;\;&x^-=x-t,\nonumber\\
&\partial_+\equiv\partial /\partial x^+,&\partial_-\equiv\partial /\partial
x^-.
\nonumber\end{eqnarray}
In the limit $S\rightarrow 0$, equations (2.8) go to the self-duality
conditions
for the non-Abelian fields $g_L,\;g_R$. Therefore, we will call equations (2.8)
self-duality conditions. At the quantum level we will show that the conformal
points of the non-Abelian
Bosonic Thirring model with the action given by eq. (2.6) are in one-to-one
correspondence to the
conformal points of the Lefton-Righton Thirring model [13].

Classically the theories (2.1) and (2.6) are distinguished, whatever
conditions we
may impose upon them. However, at the quantum level the fermionic and bosonic
non-Abelian Thirring models become indistinguishable under the following
conditions: 1) the two Weyl spinors $\psi^i_R$ and $\psi^{\bar i}_L$ carry
flavor indices $i=1,...,k_R$ and $\bar i=1,...,k_L$ and 2) the coupling
constant
matrix $S$ is invertible. When these conditions are fulfilled the statistical
sums of the two models are identical [14]. Note that the second condition is
also necessary for the Lefton-Righton Thirring model to be written in first
order form [13]. Since we are going to use the Fermi-Bose equivalence in
what following, we recall the main steps of its proof.

It has been shown in [13] that the partition function of the Bosonic Thirring
model possesses the property
\begin{equation}
Z_B(k_L,k_R;S)=J\;Z_B(k'_L,k'_R;S')\;Z_B(k_L,k_R;0)\;Z_{gh},\end{equation}
where
\begin{eqnarray}
Z_B(k_L,k_R;S)&=&\int {\cal D}g_L{\cal D}g_R\;{\rm e}^{-S_B},\nonumber\\ & & \\
Z_{gh}&=&\int{\cal D}b{\cal D}\bar b{\cal D}c{\cal D}\bar c\exp \left[
-\int d^2z(b\bar\partial c+\bar b\partial\bar c)\right],\nonumber\end{eqnarray}
with $(b,c)$ and $(\bar b,\bar c)$ Grassmann odd auxiliary fields from the
adjoint representations of ${\cal G}_L$ and ${\cal G}_R$ respectively. The
constant $J$ in eq. (2.9) is a jacobian factor due to the change in the measure
of the auxiliary fields [2].
The
relation (2.9)
allows us to see how the Bosonic Thirring model partition function
transforms under inversions of the Thirring couplings
\begin{equation}
S_{a\bar a}\rightarrow S'_{a\bar a}=-\left(k'_Lk'_RS_{a\bar b}S_{b\bar
b}\right)
^{-1}S_{b\bar a}\end{equation}
and simultaneous mirror reflections of the central charges
\begin{equation}
k_L\rightarrow k'_L=-k_R-c_2(G_R)/2,\;\;\;\;k_R\rightarrow
k'_R=-k_L-c_2(G_L)/2,
\end{equation}
where $c_2(G_L)$ and $c_2(G_R)$ are quadratic Casimir operator eigenvalues
referring to the adjoint representations of ${\cal G}_L$ and ${\cal G}_R$
respectively.

To link the partition function of the Bosonic Thirring model with the partition
function of the Fermionic Thirring model we consider an equivalent dual
formulation of the Fermionic Thirring Lagrangian
\begin{equation}
\tilde L_F(\psi_R,\psi_L;A_+,A_-;S)=\bar\psi_L\bar\partial\psi_L + \bar\psi_R
\partial\psi_R +A_-^aJ_L^a+A_+^{\bar a}J_R^{\bar a}+(S^{-1})_{a\bar a}A_-^a
A_+^{\bar a}.\end{equation}
It is easy to see that this theory gives rise to the Fermionic Thirring model
after eliminating the auxiliary fields $A_-$ and $A_+$ by using their algebraic
equations of motion. Due to the algebraic character of the auxiliary fields,
the
dual equivalence must hold also at the quantum level. Then the Fermionic
functional integrals in the partition function
\begin{equation}
Z_F=\int {\cal D}\psi_L{\cal D}\psi_R\;{\rm e}^{-S_F}
\end{equation}
can be computed by the chiral anomalies resulting in the non-local functional
of
the auxiliary fields [15]. These non-local expressions transform to the WZNW
models after making the change
\begin{equation}
A_-\rightarrow\bar\partial g_Rg_R^{-1},\;\;\;\;A_+\rightarrow g^{-1}_L\partial
g_L.\end{equation}
Arising within the process the partition function of the Bosonic Thirring model
leads us via property (2.9) to the remarkable identity [14]
\begin{equation}
{Z_B(k_L,k_R;S)\over Z_B(k_L,k_R;0)}={Z_F(k_L,k_R;S)\over Z_F(k_L,k_R;0)}.
\end{equation}

Apparently, in the limit $S=0$ the identity (16) becomes trivial.
This is not surprising because as we demonstrated in [10] in order to
fermionize
the WZNW models (or $S=0$ Bosonic Thirring model) with arbitrary levels, we
have to use
the Fermionic Thirring model at the so-called isoscalar Dashen-Frishman
conformal points, not at $S_{a\bar a}=0$. We will discuss this procedure in
sect. 6 of the
present paper. Meanwhile, when $S\ne 0$, the identity (2.16) is very fruitful
since it allows us to establish an equivalence between the conformal points of
the
Fermionic and Bosonic versions of the Thirring model, and to clarify its
geometrical meaning [5,4]. Furthermore, we can easily show that the ratio of
the
partition functions for the non-Abelian Bosonic Thirring model is equal to
the ratio of the partition functions of the Lefton-Righton Thirring model.
Indeed the partition functions of the non-Abelian and
Lefton-Righton Thirring models differ only by the ghost partition function.
This ghost partition function
does not depend on the coupling constants. Therefore, the ghost
contributions are the same in the nominator and denominator of eq. (2.16) and,
hence, cancel each other.

It follows also from formula (2.16)
that only chargeless combinations of fermions
contribute to the normalized partition function. Therefore, to preserve the
Lorentz symmetry at the quantum level, it would be sufficient to keep it
apparent only for the mentioned composite chargeless fields. While fermions
themselves might not be of any certain spin. Later on we will discuss
how this phenomena may affect the existence of massive deformations of
conformal non-Abelian Thirring models.

\section{Lie-Poisson structure of Hamiltonian system}

Identity (2.16) will continue to be somewhat formal until we are able to
calculate the functional integrals for the non-Abelian Bosonic and Fermionic
Thirring models. Apparently, this seems to be very difficult for
arbitrary values of the coupling constant matrix $S_{a\bar a}$. However, it
might be possible at some particular values of $S_{a\bar a}$ at which the
theory could be quantized nonperturbatively.

All currently known nonperturbative quantum methods are essentially based on
some symmetries which can be promoted through Poisson brackets to the quantum
level. Therefore, the first thing we have to learn about the non-Abelian
Thirring model is to find its phase space symmetries. This will be a subject
of the present section. Specifically we will focus on the symmetries of the
non-Abelian Bosonic Thirring model.

The non-Abelian Bosonic Thirring model is a highly nonlinear field theory.
Therefore, its analysis is very complicated. However, its
symmetries can be uncovered from another simpler model which has a very nice
geometrical structure [5, 6]. This geometrical theory is described by the
following action
\begin{equation}
S_G=\int\alpha_L\;+\;\int\alpha_R\;+\;S_H,\end{equation}
where $\alpha_L$ and $\alpha_R$ are canonical one-forms associated to the
nondegenerate closed symplectic structures defined on the coadjoint orbits of
the affine groups $\hat G_L$ and $\hat G_R$ respectively [17, 18]. The
relations between $\alpha_L,\;\alpha_R$ and $\omega_L,\;\omega_R$ are given
locally by
\begin{equation}
{\rm d}\alpha_L=\omega_L,\;\;\;\;\;\;{\rm d}\alpha_R=\omega_R.\end{equation}
The last term in eq. (3.17) is defined by a Hamiltonian in the phase space with
the symplectic forms $\omega_L$ and $\omega_R$. The explicit expressions for
the symplectic forms are
\begin{eqnarray}
\omega_L&=&(k_L/\pi)\int tr_L(Lg^{-1}_Ldg_L\wedge g_L^{-1}dg_L-dL\wedge
g_L^{-1}dg_L)dx^+dx^-,\nonumber\\& &\\
\omega_R&=&(k_R/\pi)\int tr_R(Rdg_Rg_R^{-1}\wedge dg_Rg_R^{-1}-dR\wedge
dg_Rg_R^{-1})dx^+dx^-,\nonumber\end{eqnarray}
where $L$ and $R$ are fields conjugated to $g_L$ and $g_R$ respectively.

The corresponding Poisson brackets between the canonical variables are found by
inverting $\omega_L$ and $\omega_R$ [19]. We find
\begin{eqnarray}
\{g_L^1(x^+),g_L^2(y^+)\}&=&0,\nonumber\\
\{L^1(x^+),g^2_L(y^+)\}&=&-2\gamma_LC_Lg^2_L(y^+)\delta(x^+-y^+),
\;\;\;\;\;\gamma_L=\pi/k_L,\nonumber\\
\{L^1(x^+),L^2(y^+)\}&=&(\gamma_L/2)[C_L,L^1(x^+)-L^2(y^+)]\delta
(x^+-y^+)+\gamma_LC_L\delta '(x^+-y^+);\nonumber\\& &\\
\{g^1_R(x^-),g^2_R(y^-)\}&=&0,\nonumber\\
\{R^1(x^-),g^2_R(y^-)\}&=&-2\gamma_RC_Rg^2_R(y^-)\delta(x^--y^-),
\;\;\;\;\;\gamma_R=\pi/k_R,\nonumber\\
\{R^1(x^-),R^2(y^-)\}&=&(\gamma_R/2)[C_R,R^1(x^-)-R^2(y^-)]\delta
(x^--y^-)+\gamma_RC_R\delta '(x^--y^-).\nonumber
\end{eqnarray}
Here $\{A^1(x),B^2(y)\}$ denotes either the $2\dim G_L\times 2\dim G_L$ or
$2\dim G_R\times 2\dim G_R$ matrix of all Poisson brackets $A$ and $B$,
arranged in the same fashion, as in the product of matrices
\begin{eqnarray}
A^1=A\otimes I\nonumber\end{eqnarray}
and
\begin{eqnarray}
B^2=I\otimes B,\nonumber\end{eqnarray}
with $I$ the unity either in ${\cal G}_L$ or ${\cal G}_R$. $C_L$
and $C_R$ are constant matrices given by
\begin{equation}
C_L={\sum_a} t^a\otimes t^a,\;\;\;\;\;\;C_R={\sum_{\bar a}}t^{\bar a}\otimes
t^{\bar a}.\end{equation}

The dynamics in the phase space with the symplectic structures determined by
the first two terms in eq. (3.17),
is defined by the last term in the action. Let
us consider the following choice for $S_H$
\begin{equation}
S_H={1\over2}\int dx^+dx^-{\cal H}\end{equation}
with the Hamiltonian density ${\cal H}$ given by
\begin{equation}
{\cal H}=-{\pi\over\gamma_L\gamma_R}\langle S,L\otimes R\rangle
.\end{equation}
Here the symbol $\langle ,\rangle$ implies the double tracing over group
indices of the Lie groups $G_L$ and $G_R$.

Given the Hamiltonian density we find the Hamiltonian
\begin{equation}
H_L=\int dx^-{\cal H}\end{equation}
in the phase space of variables $g_L,\;L$, and
\begin{equation}
H_R=\int dx^+{\cal H}\end{equation}
in the phase space of variables $g_R,\; R$, respectively. These Hamiltonians
yield the dynamical equations
\begin{eqnarray}
\partial_-g_L+(\pi/\gamma_R)g_L(tr_R S\;R)&=&0,\nonumber\\& &\\
\partial_+g_R+(\pi/\gamma_L)(tr_LS\;L)g_R&=&0.\nonumber
\end{eqnarray}
If the coupling constant matrix $S$ is invertible, then we can solve these
equations to express $L$ and $R$ in terms of $\partial_+g_R$
and $\partial_-g_L$ respectively. Therefore, after substitution of the
expressions for $L$ and $R$ in the functional in eq. (3.17), we get an action
in
terms of
the fields $g_L$ and $g_R$ only. It turns out that this action yields the same
equations of motion as the action of the non-Abelian Bosonic Thirring model
upon using the self-duality conditions given by eqs. (2.8). In other words, the
dynamics of the constrained non-Abelian Thirring model should be similar to the
dynamics of the Hamiltonian system with the Hamiltonian density as in eq.
(3.23)
and the Poisson structure given by eqs. (3.20). Thus, the direct quantization
of
the Hamiltonian equations should provide the quantization to the starting
Lagrangian Thirring model. The conditions when such a quantization can be
carried out will be the topic of the next section.

\section{Hamiltonian quantization}

Based on the results obtained in the previous section we may quantize the
non-Abelian Bosonic Thirring model by the Hamiltonian method. The method will
work as long as the algebraic Poisson structure given by eqs. (3.20) are
consistent with the Hamiltonian equations. In attempts to promote the
classical Lie-Poisson structure to the quantum level we are giving ourselves an
account of possible quantum deformations coming in the quantum Poisson brackets
and quantum Hamiltonians. We do not know a systematical way to control
nonperturbative corrections to classical structures. However, if such
deformations come into being, they have to occur self-consistently, i.e.
the quantum corrections should not destroy the dynamical equations. Therefore,
we will assume the classical Hamiltonian equations as exact quantum ones
up to a certain normal ordering of composite operators.

We want to quantize the Hamiltonian system by promoting the classical Poisson
structure (3.17) to the quantum level. First of all, we postulate the
following quantum brackets
\begin{eqnarray}
\left[L^1(x^+),L^2(y^+)\right]&=&(\gamma_L/2)\left[C_L,L^1(x^+)-L^2(y^+)\right]
\delta(x^+-y^+)+\gamma_LC_L\delta '(x^+-y^+),\nonumber\\& &\\
\left[R^1(x^-),R^2(y^-)\right]&=&(\gamma_R/2)\left[C_R,R^1(x^-)-R^2(y^-)\right]
\delta(x^--y^-)+\gamma_RC_R\delta '(x^--y^-).\nonumber\end{eqnarray}
If there are quantum corrections, they should result in a certain
renormalization of the operators $L$ and $R$. Let us consider one obvious
quantum effect. In the conformal regime we want the operators $L$ and $R$ to be
scaling (but not necessarily Virasoro primary) operators. In turn, due to the
Poisson structure, these operators are
required to have classical canonical scaling weights. Respectively $L$ has
wight (1,0) and $R$ has weight (0,1). Then it is not hard
to show that the following equations should hold
\begin{eqnarray}
\partial_-L|0\rangle&=&null\; vector,\nonumber\\& &\\
\partial_+R|0\rangle&=&null\; vector.\nonumber\end{eqnarray}
{\it So, at the quantum level the operators $L$ and $R$ become to be
analytical}. (The given deformations of the classical Hamiltonian equations
originate from the quantum effects similar to the chiral anomaly which
modifies the classical conserved chiral current.) Hence, the quantum Poisson
brackets realize affine algebras.
Thus, the hidden classical affine symmetry [5, 13] becomes apparent at the
quantum level and we can identify the affine generators with the renormalized
operators $L$ and $R$. However, the affine symmetry is not necessarily a
symmetry of physical states. We will see in the next section that a highest
weight affine representation, in general, appears to be nondegenerate in
energy. Nevertheless, the classical integrability on ``the second level" (see
eqs. (4.38)) may entail the full conformal invariance as a
byproduct of the fact that the Virasoro algebra of the
conformal group belongs to the enveloping algebra of the affine algebra.

The important point to be made is that eqs. (4.28) can be thought of as another
pair of Hamiltonian equations. It is interesting that at the quantum level in
the conformal regime the r.h.s. of the
classical Hamiltonian equations reduces to null vectors.

As a consequence of the Lie algebra structure of the brackets given by eqs.
(4.27) there are only two different ways of fixing the quantum brackets between
$g_L,\;g_R$ and $L,\;R$ consistently with the corresponding Jacobi identities.
Namely, the first one is to keep the classical structures as in eqs. (3.20).
The second one is to admit an exchange of representations for $g_L$ and $g_R$
as
follows
\begin{eqnarray}
\left[L^1(x^+),g^2_R(y^+)\right]&=&-2\gamma_LC_Lg_R^2(y^+)\delta (x^+-y^+),
\nonumber\\& &\\
\left[R^1(x^-),g^2_L(y^-)\right]&=&-2\gamma_RC_Rg_L^2(y^-)\delta (x^--y^-),
\nonumber\end{eqnarray}
The second choice works when $G_L=G_R=G$. More generally one can consider some
embedding of one Lie group into another.

Both options are equally good on the basis of
symmetry arguments. So, we have to turn to the dynamical equations to make a
choice.
Obviously, the Hamiltonian equations (3.26), (4.28) in the conformal regime
should admit the following representation
\begin{eqnarray}
\left[L_{-1},L\right]&=&0,\nonumber\\
\left[L_{-1},g_R\right]&=&-(\pi/\gamma_L)tr_L:S\;L\;g_R:;\nonumber\\& &\\
\left[\bar L_{-1},R\right]&=&0,\nonumber\\
\left[\bar L_{-1},g_L\right]&=&-(\pi/\gamma_R)tr_R:g_L\;S\;R:,\nonumber
\end{eqnarray}
where we defined normal ordering by the rule
\begin{eqnarray}
:L\;g_R:(z)&=&\oint{dw\over 2\pi i}{L(w)g_R(z)\over w-z},\nonumber\\& &\\
:g_L\;R:(\bar z)&=&\oint{d\bar w\over 2\pi i}{R(\bar w)g_L(\bar z)\over \bar w-
\bar z},\nonumber\end{eqnarray}
with products $L(w)g_R(z)$ and $g_L(\bar z)R(\bar w)$ being understood as
T-ordered operator product expansions [20]. The operators $L_{-1}$ and $\bar
L_{-1}$ are the generators of translations. By definition
\begin{equation}
L_{-1}=\oint {dz\over 2\pi i}T(z),\;\;\;\;\;\;\bar L_{-1}=\oint {d\bar z\over
2\pi i}\bar T(\bar z),\end{equation}
with $T(z)$ and $\bar T(\bar z)$ the holomorphic and antiholomorphic components
of the energy-momentum tensor of the conformal Thirring model under
consideration. It is interesting to point out that eqs. (4.30) appear to be a
quantum field theoretic generalization of the isotropic rotator which possesses
quantum group symmetries [30]. Therefore, it might be possible to look for
integrable deformations of the Bosonic Thirring model following in the manner
of
ref. [30].

So, what we have to do is to find the operators $T$ and $\bar T$ which produce
the r.h.s. of eqs. (4.30). With the first Lie-Poisson structure as in eqs.
(3.20)
we fail to get any expressions for $T$ and $\bar T$ obeying the Hamiltonian
equations. While the second structure given by eqs. (4.29) admits solutions for
the operators
$T$ and $\bar T$ to exist. It is easy to check that the following operators
\begin{eqnarray}
T(z)&=&L_{ab}:\tilde L^a\tilde L^b:,\;\;\;\;\;\tilde L=k_LL;
\nonumber\\& &\\
\bar T(\bar z)&=&L_{ab}:\tilde R^a\tilde R^b:,\;\;\;\;\;\tilde R=k_RR
\nonumber\end{eqnarray}
satisfy the Hamiltonian equations (4.32) provided that\footnote{The Hamiltonian
associated with the energy-momentum tensor in (4.33) can be viewed as a
Hamiltonian of the Siegel invariant
Lefton-Righton Thirring model [3] taken in the
Floreanini-Jackiw gauge (R. Floreanini and R. Jackiw, Phys. Rev. Lett. {\bf 59}
(1987) 1873). As a matter of fact we deal with the same Hamiltonian system.}
\begin{equation}
L_{ab}=S_{ab}.
\end{equation}

However, this is not the whole story. The operators in eqs. (4.33)
should form two
copies of the Virasoro algebra. Otherwise, they will not make sense of the
components of the conformal energy-momentum tensor. We can prove that the
operators $T$ and $\bar T$ give rise to the Virasoro algebras, if and only if
the matrix $L_{ab}$ satisfies the Virasoro master equation [7]
\begin{equation}
L_{ab}=2L_{ac}G^{cd}L_{db}-L_{cd}L_{ef}f^{ce}_af^{df}_b-L_{cd}f^{ce}_ff^{df}_a
L_{be}-L_{cd}f^{ce}_ff^{df}_bL_{ae},\end{equation}
with $f^{ab}_c$ and $G^{ab}$ respectively the structure constants and general
Killing metric of the Lie algebra ${\cal G}$ and Lie group $G$ respectively.
The Virasoro master equation may have many solutions [7, 25]. The entire space
of solutions possesses the symmetry under transformations from the diagonal of
$G_L\times G_R$. This comes transparently from the affine-Virasoro construction
itself and also can be verified with the invariance of the master equation
under the following transformations
\begin{eqnarray}
L_{ab}\rightarrow L_{ab}+x_h(f^{hk}_aL_{kb}+f^{hk}_bL_{ak})+{\cal O}^2(x),
\nonumber\end{eqnarray}
where $x_h$ are the infinitesimal parameters. Accordingly, the conformal
symmetry of the non-Abelian Thirring model is held on the orbits which are
built by acting with the global diagonal group on the
``physically not equivalent" solutions of the master equation. In section 6 we
will show that components of the energy-momentum tensor in the form given
by eqs. (4.33) appear naturally in the course of coupling the Bosonic Thirring
model to the 2D gravity.

Thus, there is a one-to-one correspondence between solutions of the Virasoro
master equation and the conformal points of the Bosonic non-Abelian Thirring
model. Indeed, given a solution of the master equation we can define a value of
the Thirring coupling constant at which the theory can be quantized in a
fashion consistent with the conformal invariance. The given conformal points we
will call Dashen-Frishman conformal points [10]. In the process the conformal
symmetry emerges not
as a byproduct of the Hamiltonian quantization but as its essential
ingredient.

At the same time the conformal points may tell us something about the
model beyond the conformal regime. Going off the conformal regime means giving
mass to the fundamental fields. However, massive terms can be consistent
with the
Lorentz symmetry only if the massive fields have a certain Lorentz spin.
Obviously,
if the fields do not have the correct spin at the conformal points, they
cannot get it in the vicinity of the conformal points. In such a case
there may not be a smooth way away from the conformal phase to the
massive phase. In
other words, the theory may turn out to be well defined only at the conformal
points. On the other hand,
it may happen that some of the fields do have the correct spin.
For example, some of the components of an affine group multiplet may have a
right spin. Then the theory can be moved out of the conformal points.
We will show in next section that in Thirring models at the
Dashen-Frishman conformal points there is always at least one fundamental
field with the correct Lorentz spin.

\section{Conformal non-Abelian Fermionic Thirring model}

We now go on to discuss conformal points of the non-Abelian Fermionic Thirring
model described by the action in eq. (2.1). In this case the theory is already
first order and its quantization is more straightforward compared to
the Bosonic version. The classical Lagrangian yields the following equation of
motion for $\psi_L$
\begin{equation}
\bar\partial\psi_L=S_{a\bar a}t^aJ_R^{\bar a}\psi_L\end{equation}
and a similar one for $\psi_R$. At the quantum level this equation makes
sense provided the normal ordering of its r.h.s. exists.

The classical equations of motion entail the following relations
\begin{eqnarray}
\bar\partial J_L^a&=&if^{ab}_cS_{b\bar a}J_L^cJ_R^{\bar a},\nonumber\\& &\\
\partial J_R^{\bar a}&=&i f^{\bar a\bar b}_{\bar c}S_{a\bar b}J_L^aJ_R^{\bar
c},\nonumber\end{eqnarray}
where the currents $J_L,\;J_R$ are given by eqs. (2.2). Since the Fermionic
current-current interaction
does not contain a time derivative, the fields $J_L,\;J_R$ form a standard
current Poisson algebra [12] similar to (3.20). Eqs. (5.37) are rather
reminiscent
of the Lax pair
representations of integrable systems. Nevertheless, we cannot associate a
curl free local current to the given system, with the exception of the
isoscalar case $S_{a\bar a}=\lambda\delta_{a\bar a}$. In the isoscalar case the
system of eqs. (5.37) can be presented as a zero curvature condition for a
conserved local current.

It is not difficult, however, to find a completely integrable subsystem of the
currents. This is formulated in terms of the fields
\begin{equation}
X=G_{ab}J_L^aJ_L^b,\;\;\;\;\;\;\;\bar X=G_{\bar a\bar b}J_R^{\bar a}J_R^{\bar
b}.\end{equation}
which obviously satisfy the analytisity conditions
\begin{equation}
\bar\partial X=0,\;\;\;\;\;\;\;\partial\bar X=0.\end{equation}
Thus, nonanalytic parts of the currents $J_L,\;J_R$ are irrelevant in analyzing
the system of eqs. (5.38) and (5.39).

{}From now on we will be interested only in the conformal regime of the model
in
question. It means that its quantum energy-momentum tensor should consist of
holomorphic $T$ and antiholomorphic $\bar T$ components forming two copies of
the Virasoro algebra. In order to elucidate the expressions for $T$ and
$\bar T$, we have to carefully investigate the symmetry and dynamics of the
theory under consideration.

The current algebra is consistent with the scale symmetry as long as the
Fermionic currents are the
scaling fields of the canonical weights. At the quantum
level we can prove that the scaling properties result in analyticity
conditions for the current operators. These equations are very similar to eqs.
(4.28). Thus, in the
quantum regime the starting current algebra transforms to two
copies of the affine algebra of the composite operators $J_L$ and $J_R$.
Hence, these operators can be treated as generators of local symmetries and we
can require the following local commutation relations [21, 12]
\begin{eqnarray}
\left[J_L(z),\psi(w,\bar w)\right]&=&(a+\bar a\gamma_5)\psi(w,\bar
w)\delta(z-w),\nonumber\\ & & \\
\left[J_L^a(z),\psi(w,\bar
w)\right]&=&-{1\over2}(1+\delta\gamma_5)t^a\psi(w,\bar
w)\delta(z-w)\nonumber\end{eqnarray}
and the similar ones for $J_R$ and $J_R^a$. Here $J_L,\;J_R$ are the $U(1)$
currents given by eqs. (2.4). These currents can always be defined when we are
dealing with the complex fermions. The parameters $a$ and $\bar a$
in eq. (5.40)
are not fixed by the symmetry until we demand certain requirements for the spin
of $\psi$. The constant $\delta$ in turn must be either +1 or $-1$ from the
Jacoby identity. By a
redefinition of the field $\psi\rightarrow\gamma_1\psi$, we
can always
choose $\delta$ to be, say, +1. However, this will entail a change of the
Lorentz
representation of $\psi$ in the equations of motion. Therefore, it is more
convenient to consider the commutators with different $\delta$'s. The sign of
$\delta$ cannot be fixed from the symmetry and, therefore, should depend on
the dynamics.

The commutation relations in eqs. (5.40) reflect a simple fact that the field
$\psi$ is an affine primary field. This, however, does not imply the field
$\psi$ is a conformal primary one. Note that until now all known conformal
field theories have Virasoro primary as their fundamental fields. In the
case of the non-Abelian Thirring models, we are dealing with a more general
situation.

The commutators in eqs. (5.40) provide us with the operator product expansions
\begin{eqnarray}
J_L(z)\psi(w,\bar w)&=&{(a+\bar a\gamma_5)\psi(w,\bar w)\over z-w}+
reg.,\nonumber\\& &\\
J_L^a(z)\psi(w,\bar w)&=&{{1\over2}(1+\delta\gamma_5)t^a\psi(w,\bar w)\over
z-w}+reg.\nonumber\end{eqnarray}
Here the regular parts of the OPE's depend on the specific properties of the
representation $\psi$ with respect to the conformal transformations. To clarify
this point we have to construct the Virasoro generators by using the affine
currents. In general, it could be done in many ways [22]. However, in the case
under consideration the Virasoro generators have to be consistent with the
equations of motion.

To proceed we have to define normal ordering between the currents and the
fields. In sect. 4 we made use of normal ordering as in eqs. (4.31). This
prescription
can be extended to a general case of two operators $A$ and $B$ when one of them
is analytic (holomorphic in the case at hand). So, we define normal ordering
between $A$ and $B$ in the following fashion
\begin{equation}
:A\;B:(z)=\oint{dw\over 2\pi i}{A(w)B(z)\over w-z},\end{equation}
where as usual $A(w)B(z)$ is understood as a time-ordered OPE in the framework
of the radial quantization (after corresponding Euclideanization of
the Minkowski
space-time).

The point to be made is that the l.h.s. of eq. (5.36)
is a certain translation of
$\psi$. Hence, the equation of motion has to follow from the commutation of the
Fermionic field with the proper energy-momentum tensor. Otherwise, the theory
will be inconsistent dynamically. By definition the operators of translations
are given by the following generators of the conformal algebras
\begin{eqnarray}
\partial\rightarrow L_{-1}&=&\oint{dz\over2\pi i}T(z),\nonumber\\& &\\
\bar\partial\rightarrow \bar L_{-1}&=&\oint{d\bar z\over2\pi i}\bar T(\bar
z),\nonumber,\end{eqnarray}
with $T(z)$ and $\bar T(\bar z)$ the holomorphic and antiholomorphic components
of the energy-momentum tensor.

All in all, the quantum equation of motion should be as follows
\begin{equation}
\left[L_{-1},\psi_R\right]=S_{ab}:J_L^at^b\psi_R:.\end{equation}
We have assumed that $G_L=G_R=G$. Although one can try to consider the case
when $G_L\neq G_R$ provided one of the groups can be embedded into another.

The aim is to construct the operator $L_{-1}$ and $\bar L_{-1}$ so that they
obey all the quantum
equations of motion. It turns out that the quantum canonical brackets with
$\delta =+1$ are incompatible with the full system of equations of motion.
Therefore we have to consider $\delta=-1$. In this case it is not hard to
check that eq. (5.44) is fulfilled with $T$ given by
\begin{equation}
T=L_{ab}:J_L^aJ_L^b:+\kappa:J_LJ_L:,\end{equation}
provided that
\begin{equation}
L_{ab}=S_{ab}.\end{equation}
The last term in eq. (5.45) originates from the free Fermionic theory and
survives in the interacting model because the $U(1)$ current commutes with the
non-Abelian currents. The magnitude of the constant $\kappa$ does not affect
any
observables in the theory.

Thus, with $\delta=-1$ we are able to define all quantum operators entering the
quantum equations of motion. More conditions are required for $L_{-1}$ and
$\bar L_{-1}$ to be the translation operators. The operators $L_{-1},\;\bar
L_{-1}$ will enjoy this property, if $T$ and $\bar T$ form Virasoro
algebras. With the fact that the currents $J_L^a$ satisfy the affine algebra,
we can show that $T$ forms the Virasoro algebra, if and only if the matrix
$L_{ab}$ is a solution of the Virasoro master equation (4.35). So, the Bosonic
and Fermionic non-Abelian Thirring models share the same Dashen-Frishman
conformal points in
the full correspondence with eq. (2.16). This result implies that the
partition functions in eq. (2.16) is perhaps calculable at the
Dashen-Frishman conformal points.

In general, we do not know how to handle the partition function at the
non-perturbative Dashen-Frishman conformal points. However, at the particular
conformal points corresponding to the isoscalar case we can gain some insight.

Let us consider the simplest case when $k_L=k_R=k$. By using eqs. (2.9) and
(2.16),
we can obtain the following formula for the Fermionic partition function
\begin{equation}
Z_F(k,k;S)=J\;
Z_F(k,k;0)Z_B\left(-(k+{1\over2}c_2(G)),-(k+{1\over2}c_2(G));S'\right)
Z_{gh},
\end{equation}
where $S'$ is given by eq. (2.11).
The partition function of the Bosonic Thirring model possesses a useful
property
\begin{equation}
Z_B(k,k;S={\bf 1}/k)=Z_B^{1\over2}(k,k;0),\end{equation}
which is a direct consequence of the Polyakov-Wiegmann formula. Keeping in mind
the given property one can easily prove the following identity
\begin{equation}
Z_F(k,k;S={\bf 1}/(k+{1\over2}c_2(G)))=J\;Z_F(k,k;0)
Z^{1\over2}_B\left(-(k+{1\over2}c_2(G)),-(k+{1\over2}c_2(G));0\right)Z_{gh}.
\end{equation}
We have used the fact that the coupling constant $S'$ associated with the
coupling $S^{*}={\bf 1}/(k+{1\over2}c_2(G))$ is given by
\begin{equation}
S'= -{\bf 1}/(k+{1\over2}c_2(G)).\end{equation}
Therefore, we can use relation (5.48) to obtain eq. (5.49). The latter
signifies
that the coupling constant $S^{*}$ corresponds to the conformal point of the
Fermionic Thirring model since on the r.h.s of identity (5.49) we have a
product
of the conformal partition functions. The given conformal point is nothing but
the isoscalar Dashen-Frishman conformal point [12] generalized to the case of
spinors with $k$ flavors. Note that
the presented
proof of the conformal symmetry of the Fermionic Thirring model at the
isoscalar Dashen-Frishman fixed point is essentially nonperturbative.

At the same time, it may be instructive to check the conformal symmetry by
another method.
Namely, one can use the $1/N$ method.
Indeed, let us consider the $SU(N)$ non-Abelian Fermionic Thirring model with
the isoscalar current-current interaction
\begin{equation}
S_{int}=-\lambda\int d^2z\; J^a_LJ^a_R,\end{equation}
with $\lambda$ being a coupling constant. When the fermions do not have flavor,
we should be able to treat this theory by the $1/N$-expansion method in the
limit of large $N$.

An isoscalar solution of the Virasoro master equation (4.35) in the case
under consideration reads [12]
\begin{equation}
\lambda=4\pi/(N+1).\end{equation}
In the limit when $N$ is large the following ratio
\begin{equation}
4\pi N\;\lambda=\bar\lambda=1\end{equation}
holds. Hence, one can use the $1/N$-expansion method to explore the theory at
the given value of the coupling constant.
Obviously, if the model had conformal points at
nontrivial values of $\lambda$, then the corresponding renormalization group
$\beta$-function should vanish at these points for each order in $1/N$.
Actually, one can use the results
obtained in [23, 24]. It is known that the
isoscalar Thirring model is equivalent to the Gross-Neveu theory for a small
coupling [14]. A puzzle is that Gross and Neveu have shown that their model
does not allow
nontrivial conformal points to exist. At the same time, we just proved above
that the Dashen-Frishman model does have a nontrivial conformal point. An
explanation of the paradox might be as follows.  In the small vicinity of zero
coupling constant there may exist a phase
transition which prohibits Fierz transformations to be used at the values of
$\lambda$ comparable with the value given by eq. (5.52).
Therefore, the nontrivial
conformal point of the non-Abelian Thirring model can be missed in the
Gross-Neveu theory. Our conjecture is that the Gross-Neveu model describes the
non-Abelian Thirring model in the
phase of very small couplings. Whereas  when $\lambda$ approaches
the critical value from the right side on a parametrical line the
Dashen-Frishman isoscalar model seems to be equivalent to
the Wilson's theory of $N$ scalar fields
in $d=4-\epsilon$ dimensions in the limit $\epsilon\rightarrow 2$. Wilson has
calculated the dependence of
the coupling $\lambda$ on the cut-off in the given limit and he established
that the model has a nontrivial fixed point
when $\bar\lambda=1$ (in our normalization). Thus, the conformal point in eq.
(5.52) which follows from the Hamiltonian quantization could be in a favorable
agreement
with the conformal point in eq. (5.53)
following from the $1/N$-expansion method.

Of course, the $1/N$-expansion method fails to be appropriate for most of other
Dashen-Frishman conformal points. However, one could try to apply it to the
case when the number of colors is fixed but the number of flavors goes to
infinity. This situation corresponds to the case of the affine algebra with a
large level.

{}From the point of view of the Hamiltonian quantization, the consistency of a
solution of the Virasoro master equation with the conformal invariance of
the quantum field theory
is enough to justify the conformal symmetry of the non-Abelian
Thirring models at all other Dashen-Frishman conformal points.

It is noteworthy that the non-Abelian Fermionic Thirring model at the isoscalar
conformal points corresponding to the affine-Sugawara construction [7] yields
the
proper Fermionic Lagrangian description of the representation described by the
WZNW model on affine $\hat G$. Thus, in order to fermionize the WZNW model with
the level permitted by the bifermionic currents, we have to take fermions
described not by the free Lagrangian but the isoscalar Thirring Lagrangian.

Now we would like to discuss some features of the fundamental fields at the
Dashen-Frishman conformal points. We begin with the vacuum of affine $G$
\begin{equation}
J^a_{m\ge 0}|0\rangle=L_{m\ge-1}|0\rangle=0,\end{equation}
where $J_m^a$ and $L_m$ are defined as
\begin{eqnarray}
J_L^a(z)&=&{\sum^\infty_{m=-\infty}}J_m^az^{-m-1},\;\;\;a=1,...,\dim
G,\nonumber\\& &\\
L_{ab}:J_L^a(z)J_L^b(z):&=&{\sum_{m=-\infty}^\infty}L_mz^{-m-2}.\nonumber
\end{eqnarray}
Due to the property in eq. (5.41), the field $\psi_R^\alpha$ has to obey
\begin{equation}
J^a_{m\ge0}\psi_R^\alpha(0)|0\rangle=\delta_{m,0}\left(t^a\right)^\alpha_\beta
\psi^\beta_R(0)|0\rangle.\end{equation}
Consider now
the action of the $L_{ab}:J_L^aJ_L^b:$ on the affine primary states.
It is
easily verified with eq. (5.54) that
\begin{equation}
L_0\psi_R^\alpha(0)|0\rangle=\Delta^\alpha_\beta\psi^\beta_R|0\rangle,
\end{equation}
where
\begin{equation}
\Delta^\alpha_\beta=L_{ab}\left(t^at^b\right)^\alpha_\beta\end{equation}
is called the conformal weight matrix [25]. There exists an eigenbasis of
affine primary fields in which
the conformal weight matrix is diagonal [26]. When the
field $\psi$ is in the fundamental representation,  we
may think of the fundamental fields $\psi^\alpha$ as the eigenbasis of the
affine primary fields.

The information about the Lorentz spin of $\psi_R$ resides in the matrix
$\Delta_\alpha^\beta$. Namely, the Lorentz spin operator is given by
\begin{equation}
s=L_0-\bar L_0.\end{equation}
Since in the $\{\psi^\alpha\}$
eigenbasis the conformal weight matrix has a diagonal form,
the eigenmatrix of the spin
operator also can be arranged to be diagonal. Thus, in general, different
components of an affine multiplet should be of a different spin. Due to this
fact, the components of
an affine multiplet describe different conformal highest weight
representations with different background energies. Therefore, the underlying
affine symmetries
are not generally symmetries of the physical states, since the affine
generators may not commute with the Hamiltonian.\footnote{In the
appendix we will
discuss the situation when all components of an affine
multiplet have the same spin such that the affine field becomes simultaneously
a Virasoro primary.}

Since the matrix $L_{ab}$ is fixed by the conformal symmetry, we cannot
change the eigenvalues of $s$ by tuning $L_{ab}$. However, there is one free
parameter - the parameter $a\;(\bar a=a$ in eq. (5.40) when the $U(1)$
current-current interaction is omitted in the starting Lagrangian).
Via the operator
$\bar L_0$ this parameter enters the eigenmatrix of $s$. This enables us to set
the spin of one of the components of an affine multiplet to any value we want.
This is a consequence of the fact that the
diagonal of the $U(1)\times U_{\gamma_5}(1)$
symmetry is conserved at the quantum level at arbitrary values of the Thirring
coupling constants. Therefore, in
the case of the Fermionic theory, both at the classical and quantum levels
one of the fermions may have Lorentz spin equal to 1/2 .
This is very important, if one wants to consider the Thirring
model beyond conformal points. Indeed, the existence of the fundamental fields
of undeformed Lorentz spin allows massive terms to appear.
This may also mean that
the space of all Thirring models is a connected multitude.

\section{A BRST nature of the master equation}

In this section we are going to show that the Bosonic Thirring model can be
viewed as a gauge invariant theory such that the action in eq. (2.6)
corresponds to a particular gauge choice in the gauge model. Interestingly,
one gauge symmetry comes into being due to the chirality conditions given by
eqs. (2.18). This is the Siegel gauge symmetry [31]
arising in the process of including the chirality constraints in the
Lagrangian. This symmetry is anomalous at the quantum level. However, there is
a remedy to cure this problem [32, 33]. Due to Siegel's symmetry all auxiliary
lagrange multipliers become pure gauge degrees of freedom and can be set to
zero
values both at the classical and quantum levels [32, 33].
Presently we will be convinced of an
importance of the given symmetry
for understanding the equivalence between the Bosonic Thirring model and the
affine-Virasoro construction within the BRST approach. In order to see more
local symmetries, we should look at the
global symmetries of the Bosonic Thirring model.

Let us forget for a while about the interaction term in eq. (2.6). Then each of
the two WZNW models should possess global symmetries generated by the following
conserved currents
\begin{equation}
{\cal J}^{a_1a_2...a_n}_n=P^{a_1a_2...a_n}_n(J,\partial),\end{equation}
where $P_n$ is a polynom of order $n$ in $J^a$ and $\partial$ with $J^a$
being the affine
current obeying the equation of motion
\begin{equation}
\bar\partial J^a=0.\end{equation}
The currents presented in eq. (6.60) may form a very rich algebra. In this
paper
we want to concentrate on a particular subalgebra of this big algebra formed by
the currents ${\cal J}^a_1=J^a$ and
\begin{equation}
T=L_{ab}{\cal J}_2^{ab}=L_{ab}J^aJ^b
\end{equation}
without $\partial$-dependent terms. The current $T$ carries spin two and is a
natural candidate on a role of the energy-momentum tensor. There is also an
antiholomorphic component
\begin{equation}
\bar T=L_{ab}\bar J^a\bar J^b.\end{equation}
Note that in general the current $\bar T$ may go with a different matrix $\bar
L_{ab}$. However in what follows we will be restricted to the case when $\bar
L_{ab}=L_{ab}$.

The very important point to be made is that the classical currents $T$ and
$\bar T$ form a closed algebra provided the matrix $L_{ab}$ obeys the following
algebraic equation
\begin{equation}
L_{ab}=2L_{ac}G^{cd}L_{db},\end{equation}
which is easily identified with the classical limit of the Virasoro master
equation [8].

Correspondingly, transformation properties of fields follow from the
formulas
\begin{eqnarray}
\delta_T\psi&=&\epsilon\oint{dw\over2\pi i}T(w)\psi,\nonumber\\ & & \\
\delta_{\bar T}\psi&=&\bar\epsilon\oint{d\bar w\over2\pi i}\bar T(\bar w)\psi,
\nonumber\end{eqnarray}
where $\psi$ is a field and the product on the r.h.s is
understood as an OPE; $\epsilon,\;\bar\epsilon$ are constant parameters. The
given definition is suitable for
WZNW models since group elements are affine primary fields whose OPE's
with the affine currents are known.

In order to gauge the symmetry associated with the conserved currents $T$ and
$\bar T$ within the WZNW model,
one has to introduce a set of new fields $h,\;\bar h$ coupling to the
group element trough the currents $T$ and $\bar T$ respectively. The procedure
is rather straightforward and to a great extent is reminiscent of the method
[34] used to construct the gauge theory of the $W$-gravity. We are not going
into all details of this method.
For us it is
important to point out that the gauge fields $h,\;\bar h$ can be identified
with a metric of the 2D gravity [34].
Thus, we can conclude that the 2D gravity can
couple to the classical
WZNW model in as many ways as a number of solutions of the
equation (6.64) can be found. At the quantum level one can expect to get more
restrictions on classically admitted solutions.

Let us turn to the Bosonic Thirring model. Now the theory has to describe the
interaction between two chiral
WZNW models coupling to the 2D gravity. Such an
interaction can be constructed consistently with the group of two
dimensional diffeomorphisms with a method developed in ref. [3].

The interaction term of the Bosonic Thirring model spoils
the analiticity properties of the affine currents. Therefore, the
energy-momentum tensor in general will be different from the affine-Virasoro
form. It is quite amusing that when the coupling constant matrix $S_{ab}$
coincides with the
momentum matrix $L_{ab}$, i.e.
\begin{equation}
S_{ab}=L_{ab},\end{equation}
the components of the energy-momentum tensor of the Bosonic Thirring theory
acquire the affine-Virasoro construction form
given by
\begin{equation}
T_L\equiv4\pi\left({\delta S\over\delta h}\right)_{h,\bar h=0}=L_{ab}J^a_LJ^b_L
,\;\;\;\;\;\;T_R\equiv4\pi\left({\delta S\over\delta\bar h}\right)_{h,\bar h=0}
=L_{ab}J^a_RJ^b_R,\end{equation}
where
$J_L=kg_L^{-1}\partial_+g_L\;+\;k\partial_+g_Rg_R^{-1},\;J_R=k\partial_-g_R
g_R^{-1}\;+\;kg_L^{-1}\partial_-g_L.$
To get these expressions one has to use the classical master equation (6.64).
It is quite natural that eqs. (6.67) appear to be similar to the usual
spin-spin
Hamiltonian interaction of a system of two identical rotators [30].
Obviously, $T_L$ and $T_R$ satisfy the analiticity conditions
\begin{equation}
\bar\partial T_L=0,\;\;\;\;\;\;\;\partial T_R=0.\end{equation}
So that nonanalytical parts of the currents $J_L,\;J_R$ become irrelevant in
eqs. (6.68).

At the quantum level instead of the classical $T_L$ and $T_R$ we
consider the quantum energy-momentum tensors with corresponding
ghost contributions. Given the quantum $T_L$ and $T_R$ one can construct the
BRST operator $Q$ following the standard scheme [35]. It is well known that
the nilpotence of $Q$
guarantees the conformal symmetry of the system. In its turn the conformal
symmetry entails the analiticity of the quantum currents $J_L$ and $J_R$. After
that we come to the affine-Virasoro construction and the Virasoro master
equation. However, now we have one more restriction coming from the nilpotence
of the BRST operator. Namely,
\begin{equation}
c=26,\end{equation}
where $c$ is the Virasoro central charge of the affine-Virasoro construction
\begin{equation}
c=2G^{ab}L_{ab}.\end{equation}

The last restriction is
not very severe, since the Bosonic Thirring model is
to be considered as a conformal model describing a compactification of a
certain
string theory. Therefore, in a whole theory the nilpotence of $Q$ will result
in
the condition
\begin{equation}
c_0\;+\;c=26,\end{equation}
where $c_0$ is a total Virasoro central charge of a noncompact part of a given
string.

Thus, we have proved that equations (4.34), (4.35), (6.66) appear to be
necessary and
sufficient conditions of the conformal invariance of the Bosonic Thirring
model. Due to the identity (2.16), the Fermionic Thirring model should  share
the same
conformal conditions.

Note that in ref. [8] authors discussed a gauge invariant action for the
affine-Virasoro construction by utilizing one WZNW model. We found that it is
more convenient to consider two interacting chiral WZNW theories since the
chirality conditions taken
together with the gravity constraints fix completely the
so-called K-conjugate invariance  of the affine-Virasoro construction [7, 8].
Moreover, this way we were able to discover new class of conformal quantum
field models which could be useful for description of new string
compactifications\footnote{A possibility to derive the Virasoro master equation
as a condition of conformal invariance of a certain sigma model was discussed
in [36]. However, obtained results do not seem to go beyond a classical limit
of
the master equation. To such an extent of accuracy, there are no
contradictions between our
approach and the method of beta functions in [36].}

\section{Conclusion}

Following observations of the Lie-Poisson structure and the existence of
nontrivial conformal points in non-Abelian Thirring models [5, 6, 10], we have
derived a theory of conformal non-Abelian Thirring models both for Bosonic and
Fermionic versions. We have shown that these models can be quantized in a
conformally invariant fashion at the values of the Thirring coupling constants
which are solutions to the Virasoro master equation. Due to this fact, the
conformal non-Abelian Thirring models seem to
provide the algebraic affine-Virasoro construction
with a natural Lagrangian description. Moreover, since Thirring models have a
nice interpretation in string theory [2, 3], we can expect to get explicit
conformal sigma models corresponding to the affine-Virasoro construction.

In this paper, we discussed non-Abelian Thirring models only in the conformal
regime. However, for the realization of the background independent string field
theory formulation program it is very important to investigate Thirring models
beyond conformal points. We argued that massive deformations are not
prohibited by the Lorentz symmetry even though
the quantum fermions, in general,
are no longer of Lorentz spin 1/2. Therefore, we hope that the whole
multitude of Thirring models can be realized as a connected space of theories.

\par \noindent
{\bf Acknowledgement}

I would like to thank J. M. Figueroa-O'Farrill, J. Gates, M. Green,
M. Halpern, C. Hull, N. Obers, M. Porrati,
E. Ramos, K. G. Selivanov and A. Turbiner for fruitful discussions. I thank
Chris Hull for enlightening discussions on a BRST nature of the master
equation.
I thank also J.U.H. Petersen for the careful reading of the manuscript.

{\large\bf Appendix}\vspace{.15in}
\renewcommand{\theequation}{A.\arabic{equation}}
\setcounter{equation}{0}

The peculiar algebraic properties of the affine-Virasoro construction entail
unwanted conformal representations of the Virasoro algebra. Some general
features of the representations of the affine-Virasoro constructions on a
general affine $\hat G$ can be found in [25-28]. Generally since the
affine-Virasoro construction is made with the affine currents it would
seem to be true that Virasoro representations carried by the affine-Virasoro
construction should be described in terms of the representations of the
underlying affine algebra. Such a situation occurs, for example, in the
WZNW model [29]. However, a consideration of the generic affine-Virasoro
construction makes this intuition not so obvious [26]. The
complications are partly caused by the fact that since we are dealing with the
affine-Virasoro construction, affine primary fields are no longer Virasoro
primary fields in general [27].

Surprisingly, the affine-Virasoro construction on the affine $SU(2)$ does
appear an exception to the rule. This comes about due to the following
fortunate quaternionic identity
\begin{equation}
2t^at^b=\eta^{ab}+if^{ab}_ct^c,\end{equation}
which holds when $t^a$ are the $SU(2)$ generators in the fundamental
representation of the Lie algebra
$SU(2)$. In this case, the conformal weight matrix
$\Delta^\alpha_\beta$ takes the form
\begin{equation}
\Delta^\alpha_\beta=(c/4k)\delta^\alpha_\beta,\end{equation}
where the constant $c$ is the Virasoro central charge in the $SU(2)$
affine-Virasoro construction
\begin{equation}
c=2G^{ab}L_{ab}.\end{equation}
In this case the Lorentz spin of $\psi$ is given by
\begin{equation}
s=c-4\kappa a^2.\end{equation}
Therefore, we are able to treat the field $\psi$ as a spinor when
\begin{equation}
4a^2\kappa=c-1/2.\end{equation}
Eqs. (2) and (4) remain true for all the possible solutions of the Virasoro
master
equation of the generic $SU(2)$ affine-Virasoro construction when the latter
acts on the space of states in the fundamental representation of the Lie
algebra $su(2)$. This means that the $SU(2)$ affine primary fields, say
$\psi^\alpha$, can be also considered as the Virasoro primary fields provided
the matrix $L_{ab}$ obeys the Virasoro master equation.

Keeping in mind the identity (2), we obtain the OPE between the $SU(2)$ affine
Virasoro construction $T=L_{ab}:J^aJ^b:$ and the affine primary field $\psi$
\begin{equation}
T(z)\psi^\alpha(w,\bar
w)=\Delta\left({1\over(z-w)^2}+{\partial\over\Delta(z-w)}\right)\psi^\alpha(w,
\bar w)+reg.\end{equation}
where
\begin{eqnarray}
\Delta&=&c/4k,\nonumber\\& &\\
J^a(z)\psi^\alpha(w,\bar
w)&=&(t^a)^\alpha_\beta\left({1\over(z-w)}\psi^\beta(w,\bar w)+{1\over
2\Delta}\partial\psi^\beta(w,\bar w)\right)+{\cal
O}(z-w).\nonumber\end{eqnarray}

Let us consider the following composite field
\begin{equation}
S_{ab}(t^a)^\beta_\alpha\psi^{\alpha a}=\partial\psi_R^\beta
-S_{ab}(t^a)^\beta_\alpha
:J^a\psi_R^\alpha:.\end{equation}
This field might appear in the process of quantization of the classical
equation of motion of the non-Abelian Fermionic Thirring model
\begin{equation}
\partial\psi^\alpha_R=S_{ab}(t^b)^\alpha_\beta J^a\psi^\beta_R .\end{equation}
The quantization will be consistent provided the l.h.s. of eq. (8) is a null
vector. By using the OPE's in eqs. (6), (7), one can derive the correlator of
the given composite fields [26]
\begin{equation}
\langle \psi^\alpha_a({\cal T},z)\psi^\beta_b(\bar{\cal T},
w)\rangle={\left(G_{ab}+{2\Delta-1\over
2\Delta}t_bt_a-t_at_b\right)^{\alpha\beta}\over(z-w)^{2\Delta+2}},\end{equation}
where ${\cal T}$ refers to the fundamental representation of $G=SU(2)$. Then,
by straightforward calculation, one can obtain the explicit
expression for the correlator
\begin{eqnarray}
K^{\alpha\beta}&=&\langle L_{ab}\psi^{\gamma a}(t,z)(t^b)^\alpha_\gamma
L_{cd}\psi^{\sigma c}(\bar t, w)(\bar t^d)^\beta_\sigma\rangle\nonumber\\& &\\
&=&{1/2\over(z-w)^{2\Delta +2}}\left(L_{ab}(t^at^b)^{\beta\alpha}-(1/\Delta)
L_{ab}L_{cd}(t^dt^ct^at^b)^{\beta\alpha}\right).\nonumber\end{eqnarray}
Here $\bar t$ is the complex conjugate representation defined as
\begin{equation}
(\bar
t^a)^\beta_\alpha=-\eta_{\alpha\gamma}\eta^{\beta\sigma}(t^a)^\gamma_\sigma,
\end{equation}
where $\eta_{\alpha\beta}$ is the metric which is used to rise and lower
indices
of the field $\psi$.
Now it is clear that this correlator vanishes when $t^a$ are in the
fundamental representation of $SU(2)$.

Thus, the l.h.s of eq. (8) is nothing but a null vector. Note that in order to
derive eq. (11) we have to use the Virasoro master
equation. Therefore, the vanishing of the correlator in eq. (11) can be
considered as another way for the master equation to be arrived.

\end{document}